\documentclass[12pt]{article} 
\baselineskip

\usepackage{graphicx}
\usepackage{epsfig}

\begin{document}

\begin{center}
{\bf{Ideal MHD Ballooning modes, shear flow and the stable continuum}}\\
{\em{ J B Taylor}}\\
Radwinter, Wallingford, OX10 9EJ,  UK \\
June 2012
\end{center}

\begin{flushleft}

Abstract \newline
There is a well established theory of  Ballooning modes in a toroidal plasma.   The cornerstone of this is a local eigenvalue $\lambda $ on each magnetic surface - which also depends on the Ballooning phase angle $k$.   In stationary plasmas,  $\lambda(k)$ is required only near its maximum, but in rotating plasmas its average over k is required.   Unfortunately in many cases $\lambda(k)$ does not exist for some range of $k$,  because the spectrum there contains only a stable continuum.     This limits the application of  the theory, and raises the important question of whether this ``stable interval" gives rise to significant damping.  This question is re-examined using a new, simplified, model - which leads to the conclusion that there is no appreciable damping at small shear flow.   In  particular, therefore, a small shear flow should not affect Ballooning mode stability boundaries.

\newpage
\underline{1.   Introduction} \newline
MHD Ballooning modes, that is perturbations with short wavelength perpendicular to the local magnetic field, but long wavelength  parallel to it, are a particularly insidious form of plasma instability~\cite{CHT2004}.   In an axi-symmetric toroidal plasma they are normal modes with large toroidal mode number $n$.   If the plasma is stationary (no toroidal flow) then after introducing an extended poloidal coordinate $\eta$ (which removes explicit periodicity constraints~\cite{CHT.PRS}) these modes can be represented by an eikonal that describes the short wave behavior,
\begin{equation} \xi =\xi(x,\eta)\exp inq'(x\eta + S(x))
\end{equation}
where $x$ is a minor radius coordinate, constant on each magnetic surface, and $q'(x)$ is the average magnetic shear.

In the limit $ n\rightarrow \infty $, each magnetic surface then decouples from its neighbors and has its own local growth rate $\lambda(x,k)$, given as the eigenvalue
of an ODE in $\eta$, with the surface label $x$ and the phase $k=dS/dx $ (related to the poloidal location of the mode) as
parameters~\cite{CHT.PRS},
\begin{equation}
\frac{d}{d\eta}P(\eta;x,k)\frac{d}{d\eta}\;\xi(x,\eta) + Q(\eta;x,k)\;\xi(x,\eta)
=\lambda^2(x,k)\; R(\eta;x,k)\;\xi(\eta)      \label{eq:A}
\end{equation}

The coefficients $P,Q,R,$ are defined by the plasma equilibrium and are periodic in the phase angle $k$.  The boundary condition is that $\xi$ be bounded as $|\eta| \rightarrow \infty$ and the eigenvalue $\lambda(x,k)$ is also periodic in $k$.

This local growth rate $\lambda(x,k)$ contains the information necessary to construct global Ballooning modes~\cite{CHT.PRS}.   For example, near marginal stability, the Ballooning mode growth rate is
\begin{equation}
  \Lambda = \lambda_{max} - \frac{1}{2nq'}
\left[\frac{\partial^{2}\lambda}{\partial k^{2}} \;
\frac{\partial^{2}\lambda}{\partial x^{2}}\right]^{1/2}
\end{equation}
where $ \lambda_{max} $ is the maximum of $\lambda(x,k)$  wrt both $x$ and $k$.

If there is a sheared toroidal rotation $\Omega(x)$ in the equilibrium plasma, the eikonal (1) is no longer valid.   Instead the appropriate form is~\cite{Cooper}
\begin{equation} \xi(x,\eta)\exp [inq'(x\eta + S(x) + \Omega(x)t)]
\end{equation}
In the limit $n \rightarrow \infty$, each magnetic surface again decouples from its neighbors but its time dependence is no longer purely exponential.  It is of Floquet form, given~\cite{Cooper} by the PDE
\begin{equation}
\frac{\partial}{\partial\eta}P(\eta;x,\hat{k})\frac{\partial}
{\partial\eta}\;\xi(\eta,t) + Q(\eta;x,\hat{k})\;\xi(\eta,t) = \frac{\partial}
{\partial t}R(\eta;x,\hat{k})\;\frac{\partial}{\partial t}\;\xi(\eta,t) \label{eq:AA}
\end{equation}
where $\hat{k}=k+s_v t$ and $s_v = \Omega'/q'$ is the velocity shear.

When the shear velocity is small the coefficients $P,Q,R$ evolve slowly in time and we can find an ``adiabatic" solution of Eq(\ref{eq:AA}) with exponential growth rate
$\sim \exp(\int \hat{\lambda}(x,t)dt)$, where
\begin{equation}
\frac{\partial}{\partial\eta}P(\eta;x,\hat{k})\frac{\partial}
{\partial\eta}\;\xi(\eta,t) + Q(\eta;x,\hat{k})\;\xi(\eta,t) =
\hat{\lambda}^2(x,\hat{k})\; R(\eta;x,\hat{k})\;\xi(\eta)
\end{equation}
Thus $\hat{\lambda}(x,t)$ is just the local growth rate of the {\em{static}} plasma (given by Eq(\ref{eq:A})), at a time dependent phase angle $ (k + s_v t) $,  and the average growth rate of Ballooning modes in a rotating plasma is therefore
\begin{equation}
\Lambda \sim \frac{1}{2 \pi} \oint\lambda(k) dk
\end{equation}
So we see that, even with shear flow, Ballooning instabilities can be described through the local growth rate $\lambda(x,k)$ of a static plasma.   However there is a crucial  difference in the role of $\lambda$ when shear flow is significant.  In the absence of shear flow, $\lambda$ is required only in the vicinity of its maximum wrt $k$, say at
$k_m$, whereas with shear flow it is required over the full range $0<k<2 \pi$.   This seemingly small difference introduces a fundamental difficulty:  {\em{$\lambda(x,k)$ may not be defined for all $k$}}.

This problem arises when, as is often the case, Eq(\ref{eq:A}) has an unstable eigenvalue only over a restricted range of phase angle $k$, around $k_m$: outside this range it has no discrete eigenvalues - only a stable continuum
$ \lambda = \pm \;i \omega $. Within this ``stable interval" of $k$, the solution cannot be represented by an adiabatic form with  exponential time dependence. (This depends on the highest eigenvalue being well separated from all others - a condition that is clearly violated when there is only the continuum.)  It is then not clear how $\lambda$ should be interpreted (or replaced).

Formally therefore, the applicability of Ballooning theory to plasmas with shear flow is severely limited.   This limitation is particularly apparent in determining Ballooning mode stability boundaries.  By definition these are where a mode with some particular phase $k$ first becomes unstable; however in general, modes with any other phase angle will remain stable - so that near a stability boundary we expect ballooning modes to have a significant stable interval. An important issue therefore is whether, with sheared rotation, this stable interval leads to significant damping.   If not, then although even a small rotation will reduce the growth rate of Ballooning modes, it alone cannot stabilize them - consequently the formal stability boundaries for Ballooning modes will be unchanged by small rotation.

The aim of this paper is to investigate the effect of a stable interval on Ballooning modes in a plasma with small velocity shear.   The next Section describes a model of a toroidal plasma which incorporates a stable interval and introduces previous (contradictory) attempts to deal with it.  Section 3 describes a new approach to the problem.   This is discussed in detail in Section 4.  A summary and some conclusions are given in Section 5.

\underline{2. The stable interval}\newline
There are conflicting views, leading to contradictory conclusions, on how to deal with the the stable interval.  To illustrate the problem we invoke the well known $(s,\alpha)$ model~\cite{CHT1978} of a large aspect ratio Tokamak.   ($\alpha$ is the plasma pressure gradient and
$s =rq'/q$  is the magnetic shear.)   In this model, Eq(\ref{eq:AA}) takes the specific form
\begin{equation}
\frac{\partial}{\partial\eta}(1+P^2)\frac{\partial}
{\partial\eta}\;\xi(\eta,t) + Q\;\xi(\eta,t) = \frac{\partial}
{\partial t}(1+P^2)\;\frac{\partial}{\partial t}\;\xi(\eta,t) \label{eq:B}
\end{equation}
with
\begin{equation}
P = s \eta - \alpha \sin(\eta+s_v t) \;\;\;
Q= \alpha[\cos(\eta+s_v t) + P \sin(\eta+s_v t)]
\end{equation}
When the shear velocity $s_v$ is small, Eq(\ref{eq:B}) can be reduced to the simpler ``wave Eq" form
\begin{equation}
\frac{\partial^2 \psi}{\partial \eta^2} + V(\eta,s_v t)\psi =
 \frac{\partial^2 \psi}{\partial t^2} \label{eq:C}
\end{equation}
where $\psi = \xi/\sqrt{1+ P^2} $ and the ``potential" $V$ is
\begin{equation}
V(\eta,s_v t) = -\frac{[s-\alpha\cos(\eta + s_v t)]^2}{(1+P^2)^2} +
\frac{\alpha\cos(\eta+s_v t)}{(1+P^2) }
\end{equation}
The corresponding local Eq is
\begin{equation}
\frac{\partial^2 \psi}{\partial \eta^2} + V(\eta,s_v t)\psi = \hat{\lambda}^2 \psi
 \label{eq:CC}\end{equation}
A stability diagram for the $s,\alpha$ model is shown in Fig(1).  As expected, there is a large region within the conventional ``unstable" region where most phase angles are still  stable, i.e where there is a significant stable interval in $k$.

One way to deal with the stable interval~\cite{JBT1999} is to represent the perturbation as an integral over the continuum modes.   There are difficulties in justifying this approach, but the underlying picture is that continuum modes are excited as the phase
$k+s_v t$ enters the stable interval.   The ``depth" of this excitation is proportional to $ s_v$ but the duration of the subsequent stable interval is  proportional to $ 1/s_v $ - so that phase mixing during the stable interval reduces the amplitude by a fixed amount per transit through the interval.   Averaged over the whole interval this is equivalent to a damping rate that is proportional to $\Omega'$ and
{\em{vanishes}} as $\Omega' \rightarrow 0 $.

Another approach~\cite{W and C} is based on the fact that as $\eta \rightarrow \infty$ the solution of Eq(\ref{eq:C}) will consist of outgoing and incoming, reflected, waves.   If the reflected waves can be neglected it becomes appropriate to impose an
``outgoing wave only" boundary condition.   This has no effect on eigenvalues of
Eq(\ref{eq:C}) in the unstable interval -  because if $\lambda > 0$ solutions that match to outgoing waves are those which also decay as
$\eta \rightarrow  \infty$.   However, in a stable interval the change in boundary condition has a dramatic effect - because if $\lambda < 0$ a solution that matches to an outgoing wave would otherwise diverge as $\eta\rightarrow \infty$, and would be rejected as an eigenfunction.
In fact, with the ``outgoing wave" boundary condition, Eq(\ref{eq:C})  no longer has a continuum of eigenvalues $\lambda = \pm i \omega$ in the stable interval; instead it will usually have a  {\em{discrete negative}} eigenvalue.   In this case the stable interval will introduce a significant ``wave" damping that is independent of $\Omega'$ and {\em{persists as $\Omega' \rightarrow 0$}} .

This is essentially the view taken by Waelbroek and Chen \cite{W and C}, although they also introduced a further approximation.   In essence they assumed that in Eq(\ref{eq:C}) the range of $\eta$ can be separated into an inner region where the inertial term $\lambda^2$ is negligible, and an outer region where $V$ is negligible.   One then solves
Eq(\ref{eq:C}) for $\psi$ in the inner region (with $\lambda^2=0$) and at the inner region boundary matches its logarithmic derivative $\Delta'$ to an outgoing wave.  Then $\lambda \equiv  \Delta'$ and the problem of finding an eigenvalue of the full Eq(\ref{eq:C}) with the outgoing wave boundary condition is avoided.

Whether or not one adopts this additional approximation,  (which may be valid only for  small  $\Delta'$), the crucial question is whether reflected waves {\em{can}} be neglected.   The fact that the potential $V$ is small at large $\eta$ certainly makes the reflection coefficient small.   But if the perturbation were decaying exponentially during the stable interval, as it does when reflection is ignored, then any reflected wave would have been created when the outgoing wave was exponentially larger than it is when the reflected wave returns.    This large exponential factor could outweigh the small reflection coefficient when $s_v$ is small - making the reflected wave important during the stable interval.

\underline{3. A `Toy' model}\newline
To investigate the importance of reflection we introduce a simple ``Toy" model.   Suppose first that the potential vanishes outside some small region around $\eta=0$, that is
$ V(\eta,t) = -D(s_v t) \delta(\eta) $ (so that $D$ corresponds to $\Delta' $ in the account above).
This Toy model has an {\em{exact}} solution
\begin{equation}
\psi = A \exp\int^{x-t} D(s_v t')dt'
\end{equation}
valid whether $D$ is positive (unstable interval) or negative (stable interval) and for all $s_v$.   This is precisely the result one would get from the Waelbroek and Chen approximation and appears to support their picture of the damping.   But of course the model does not yet address the question of reflections from an extended potential.   To do this we modify it by adding a weak potential ``tail" $v(\eta)$ that $\rightarrow 0 $ as
$ \eta \rightarrow \infty $.   Then $ V(\eta,t) = D(s_v t) \delta(\eta) +v(\eta)$.

At this point it is convenient to express Eq(\ref{eq:C}) (using the Green's function or more simply by introducing $ (\eta \pm t)$ as coordinates) in an integral form:
\begin{equation}
\psi(\eta,t) =
\frac{1}{2}\int^t dt' \int_{\eta-(t-t')}^{\eta+(t-t')} \psi(\eta',t') \; V(\eta',t')\; d\eta'
\label{eq:D}
\end{equation}

Then, with $V(\eta,t) = D(s_v t) \delta(\eta) +v(\eta)$, the central perturbation, $\psi(0,t) \equiv  \Psi(t) $ satisfies
\begin{equation}
\frac{d\Psi(t)}{dt} =
D(st)\Psi(t) +\frac{1}{2} \int_{0}^{\infty}\psi(\eta',t-\eta')\; v(\eta',t-\eta')\label{eq:E}
\end{equation}
Reflected waves arise from the integral term in Eq(\ref{eq:E}).   To calculate the first order reflected wave, linear in $v$, we can replace
$ \psi(\eta',t-\eta')$ in this integral by its form when $v$ is ignored - then $\psi(\eta,t)$ is constant along lines of fixed ($t-\eta$) and $\psi(\eta',t-\eta') =\Psi(t-2\eta')$.   Thus we obtain a closed Eq for the central perturbation  $\Psi(t)$.
\begin{equation}
\frac{d\Psi}{dt} =
D(st)\Psi(t) + \frac{1}{2}\int_{-\infty}^{t}\Psi(\eta') v((t-\eta')/2,(t+\eta)/2) d\eta'
\label{eq:EE}\end{equation}
(For a geometrical interpretation of Eqs(\ref{eq:D}-\ref{eq:EE}) see Figs(2,3)).

\underline{4. Calculation}\newline
The effect of a weak, but extended, potential tail can be calculated explicitly in the case that it decays exponentially $\sim \exp(-p\eta)$ and is independent of time.  Then
\begin{equation}
\frac{d\Psi(t)}{dt} =
 D(s_v t)\Psi(t) + v_0 \int_{-\infty}^{t} \Psi(\eta') \exp(-p\;(t-\eta')) d \eta'
\end{equation}
With $\tau=s_v t$, this is equivalent to the Eqs
\begin{equation}
s_v\frac{d\Psi}{d\tau} -D(\tau) \Psi(\tau) = \Phi(\tau)\;\;\; : \;\;\;
s_v\frac{d\Phi}{d\tau} + p\;\Phi(\tau) = v_0 \Psi(\tau) \label{eq:F}
\end{equation}

When $s_v$ is small, Eqs(\ref{eq:F}) have the WKB solutions
\begin{equation}
\Psi^P(\tau) =q(\tau) \; \exp(\frac{1}{s_v} \int^\tau D(\tau')d\tau') ,\hspace{1 cm}
\Psi^S(\tau) =\frac{1}{q(\tau)} \frac{1}{(D(\tau)+p)} \exp(-\frac{p}{s_v}\tau)\label{eq:K}
\end{equation}
where
\begin{equation} q(\tau) = \exp (\;(\frac{v_0}{s_v})\int^\tau \frac{1}{(D(\tau')+p)}d\tau')\label{eq:KK} \end{equation}
which, since $v_0$ is assumed small, is slowly varying compared to $\psi^P(\tau), \psi^S(\tau) $ .

Thus we see that perturbation $\Psi$ has two components.   One is essentially determined by the strength of the central potential $D$;  the other by rate of decay ({\em{not}} the strength!), of the potential tail.  For convenience we will refer to these as the Primary (P) and Secondary (S) components respectively.   The two  components are independent of each other except at the transition points where $D+p=0$, - that is near the beginning and end of a stable interval.

To describe the coupling between P and S components at the transition points it is convenient to write
$ \Psi(\tau) = \chi(\tau) \exp(-p\tau/s_v)$ . Then
\begin{equation}
s_v\frac{d^2\chi}{d\tau}- (D(\tau)+p\;)\;\frac{d \chi}{d\tau} -
(D'(\tau) +\frac{v_0}{s_v} \;)\;\chi =0
\end{equation}
At the first transition, near the start of a stable interval,  $(D+p) \sim -a\tau$ with $ a > 0 $ and
\begin{equation}
\frac{d^2\chi}{du^2} + u \frac{d\chi}{du} +(1-b)\chi = 0 \label{eq:H}
\end{equation}
where $u = \sqrt{a/s_v}\; \tau,\;\;b = v_0/ a s_v$ and the Primary and Secondary components are $\chi^P(u) = u^{-b} \exp(- u^2/2),  \;\;\;
\chi^S(u) = u^{(b - 1)} $.

The general solution of Eq(\ref{eq:H}) can be expressed in terms of Parabolic Cylinder Functions as
\begin{equation}
\chi = [A \; {\bf{D}}_{-b} (u) + B \; {\bf{D}}_{(b-1)}(iu)] \;\exp(-u^2/4)
\end{equation}
and the asymptotic expansions of the PCFs link the amplitudes of $\chi^P(u)$ and $ \chi^S(u) $ before and after the transition.  In particular, ${\bf{D}}_{-b}(u)$, links $\chi^P$, which is sub-dominant before the transition, to the combination $(\chi^P + C_1\; \chi^S) $ after the transition.   That is
\begin{equation}
\chi^P(u) \longrightarrow {\bf{D}}_{-b}(u)\exp(-u^2/4) \longrightarrow
\chi^P(u) + C_1 \chi^S(u)
\end{equation}
with the coupling coefficient $C_1 = \sqrt{2\pi}/\Gamma(b)$.

After the transition the primary component decays rapidly, so this transition effectively converts the exponentially growing perturbation in the unstable interval into a slowly varying perturbation during the stable interval.   In terms of $\tau$;
\begin{equation}
\tau^{-b} \exp(-a\tau^2/2 s_v)\longrightarrow
C_1\cdot (a/ s_v)^{(b-1/2)}\; \tau^{(b-1)}
\end{equation}

At the second transition, near the end of the stable interval,
$ D+p \sim +a \tau$ and
\begin{equation}
\frac{d^2\chi}{du^2} - u \frac{d\chi}{du} -(1+b)\chi = 0 \label{eq:J}
\end{equation}
The Primary and Secondary components at this transition are
$\chi^P(u) = u^b \; \exp(u^2/2) $ , $\chi^S = u^{-(b+1)} $, and the general solution of
(\ref{eq:J})  is
\begin{equation}
\chi = [A \; {\bf{D}}_b(iu) + B \; {\bf{D}}_{-(b+1)}(-u)]\;\exp(+u^2/4)
\end{equation}
In this case the asymptotic expansions show that $ {\bf{D}}_{-(b+1)}(u) $ links $\chi^S $, sub-dominant before the transition, to the combination $(\chi^S + C_2\;\chi^P)$ after the transition.
\begin{equation}
\chi^S(u) \longrightarrow {\bf{D}}_{-(b+1)}(u)\exp(u^2/4) \longrightarrow
(\chi^S(u) + C_2 \chi^P(u))
\end{equation}
with coupling coefficient $C_2 = \sqrt{2\pi} /\Gamma(1+b)$.
After this transition the secondary component rapidly becomes negligible compared to the exponentially growing primary component.   This transition therefore transforms the slowly varying perturbation during the stable interval back to exponential growth in the next unstable interval.   In terms of $\tau$
\begin{equation}
 \tau^{-(b+1)} \longrightarrow C_2\cdot (a/s_v)^{(b+1/2)}\; \tau^b \exp(a \tau^2/2 s_v)
\end{equation}

To calculate the full effect of passing through a stable interval we need also to determine the change in the secondary component across the interval, that is from
$ \chi^S(\tau_1) = \tau_1^{(b-1)} $ at a time $\tau_1$ immediately after the first transition, to
$\chi(\tau_2) = \tau_2^{-(b+1)}$ at a time $\tau_2$ immediately before the second transition.   From
Eqs(\ref{eq:K}, \ref{eq:KK}) this is given by
\begin{equation}
\frac{\chi^S(\tau_2)}{\chi^S(\tau_1)} =
\mu^{2 b}\;(\frac{\;\tau_2^{-(b+1)}}{\tau_1^{(b-1)}})
\end{equation}
Here $\mu$ is a numerical coefficient that depends on the precise form of $D(\tau)$ in the stable interval.   (When $ D(\tau) = -a \sin(\tau)$,  $ \mu = 2 $).

We see therefore, that if we follow the perturbation from an unstable interval, through the following stable interval and into the next unstable interval, its amplitude is changed by a factor.
 $\hat{C} = C_1 C_2\; (\mu)^{2b}\;( a/s_v)^{2b} $.

\underline{5. Summary and Conclusion} \newline
There is a well established theory of MHD Ballooning modes in a toroidal plasma~\cite{CHT2004, CHT.PRS}.   The cornerstone of this theory is a local growth rate $\lambda(x,k)$ - defined as the largest eigenvalue of an ODE
\begin{equation}
\frac{d}{d\eta}P(\eta;x,k)\frac{d}{d\eta}\;\xi(\eta) + Q(\eta;x,k)\;\xi(\eta)
=\lambda^2(x,k)\; R(\eta;x,k)\;\xi(\eta) \;\;\;\;     \label{eq:AAA}
\end{equation}
on each magnetic surface $x$. ( $\eta$ is an extended poloidal coordinate and $k$ a poloidal phase; the boundary condition is that $\xi(\eta$) be bounded as $\eta\rightarrow \infty$.)

If the plasma is at rest $\lambda(x,k)$ is required only near its maximum wrt $k$, but if the plasma has a toroidal flow, the phase $k$ increases continuously with time and $\lambda$ is required over the full range $0< k < 2\pi$.   This leads to a serious problem:  in many cases $\lambda$ is  not defined for all values of $k$.  Instead there is a ``stable interval" of $k$ where Eq(\ref{eq:AAA}) has no discrete eigenvalue - only a continuum of stable eigenvalues $\lambda = i \omega$.   The existence of this stable interval restricts the application of Ideal MHD Ballooning theory in rotating plasmas and raises the question of whether it leads to significant damping - which is particularly important when considering stability boundaries.

One suggestion~\cite{JBT1999} for dealing with a stable interval was to expand the perturbation in the (singular) continuum modes.   This led to the conclusion that the stable interval produces only a small damping that vanishes as  $\Omega' \rightarrow 0$, but it is difficult to justify this analysis.

A seemingly more satisfactory approach \cite{W and C} is based on the fact that at large $|\eta|$ solutions of Eq(\ref{eq:AAA}) consist of independent outgoing and incoming
(reflected) waves.  If the reflected waves can be neglected, an ``outgoing wave only" boundary condition becomes appropriate for Eq(\ref{eq:AAA}).  Then, in the stable interval, Eq(\ref{eq:AAA}) no longer has a continuum of stable eigenvalues: instead it may have a discrete negative eigenvalue.  In this event there would be significant damping during the stable interval - which would be independent of $\Omega'$ and persist as $\Omega' \rightarrow 0$.

The crucial question in this picture is whether reflected waves can indeed be neglected.   The analysis in this paper, based on a model that specifically considers reflections, suggests otherwise.  It appears that no matter how small, reflections {\em{determine}} the behavior during the stable interval and the negative eigenvalues of Eq(\ref{eq:AAA}) are therefore not relevant.    Any damping that does occur during the stable interval is proportional to the {\em{exponent}} in the decay of the reflection coefficient at large $|\eta|$.

In the ($ s,\alpha $) model of a Tokamak, the reflection coefficient decays only algebraically, implying negligible damping during the stable interval.  Instead its effect is felt mainly at the start and finish of that interval.   At the start there is a transition from exponential growth of the perturbation to a near constant amplitude throughout the stable interval, and at the end of the interval there is a  transition back to exponential growth. These transitions, and the stable interval, change the perturbation amplitude by a factor
$\hat{C} = (2\pi/\Gamma(b)\Gamma(1+b))\;\mu^{2b} (a/s_v)^{2b}\;$ each time the phase $k(t)$ passes through the stable interval.  (Here $ s_v/a $ determines the shear flow, $v_0/a^2 $ determines the strength of reflections and $b=v_0/a s_v $.)   At small shear
($b> 1$) this change is equivalent to an average time constant throughout the interval
\begin{equation}
\Lambda^{stable} = s_v\log\hat{C} \sim \frac{2 v_0}{a}\;(\;\log\;(\frac{2 a^2}{v_0})+1) -
\frac{1}{6}(\frac{s_v^2}{v_0}) \label{eq:Q}
\end{equation}
This result indicates that, as the shear flow $\Omega'\rightarrow 0$,~\cite{Z} this damping  is negligible compared to the growth rate in an unstable interval.   It is also negligible compared to the damping which would occur if reflections were neglected.    (In effect, therefore, the effective growth rate for Ballooning modes can be summarized as ``$ \lambda(k)$ in an unstable interval and zero in a stable interval''.)

This behavior is shown clearly in numerical solutions of the underlying model
Eqs(\ref{eq:F}).   Figs(4,5) show that the instantaneous growth rate closely follows $(\exp\int (D(\tau)/s_v)\; d\tau) $ during the unstable interval and becomes briefly negative at the start of the stable interval (where the coupling coefficient is less than unity).   It is then almost constant during the remainder of the stable interval until it returns smoothly to exponential growth at the end of the interval (where the coupling coefficient $C_2$ is large).   Figs(6,7) confirm that as $s_v$ decreases the perturbation becomes effectively constant throughout the stable interval.

These features are remarkably similar to those seen in numerical computations of Ballooning modes by Furukawa and colleagues \cite{Furu1, Furu2, Aiba}.   These computations simulate the behavior of Ballooning modes in realistic Tokamak configurations, using  initial value codes.   The example in Fig(8) (adapted from Ref.\cite{Aiba}), shows the logarithmic growth rate of the kinetic energy of an $n\rightarrow \infty $ Ballooning mode in a Tokamak with toroidal rotation.  (The ``Mach" numbers $M_i$ are proportional to the shear flow velocity.)  This closely resembles the model solutions in Figs(5, 7).

In conclusion, it appears that, contrary to some previous conclusions,  a stable interval in $k$ does not lead to significant damping of Ballooning modes in a plasma with small toroidal shear flow, and therefore does not change formal Ballooning mode stability boundaries.

I am grateful to J Connor for many valuable discussions and comments, to J Hastie for the calculation of Fig(1) and to M Furukawa and N Aiba for Fig(8).

\end{flushleft}

\begin{figure}
\centering
\includegraphics[width=\textwidth]{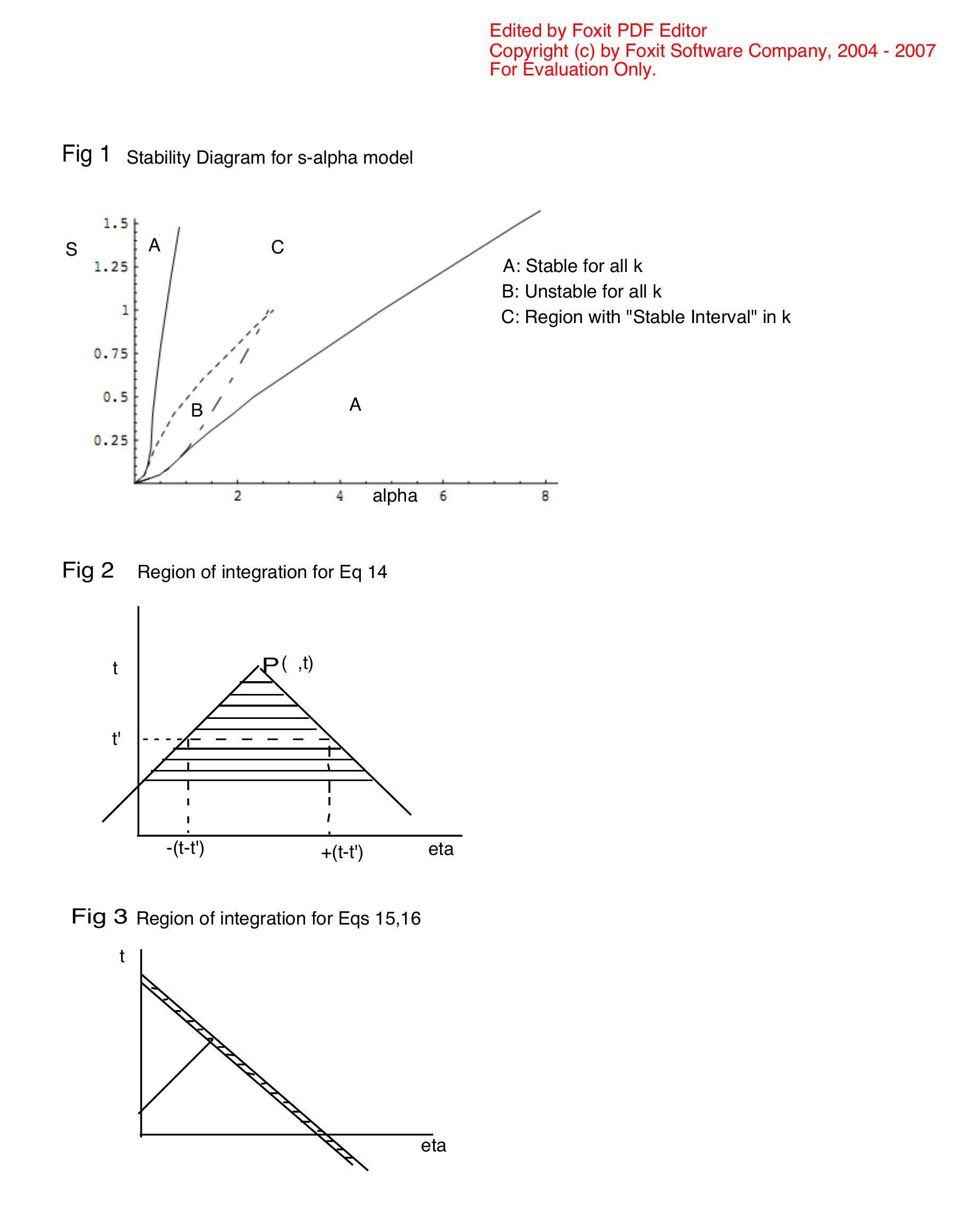}
\end{figure}

\begin{figure}
\centering
\includegraphics[width=\textwidth]{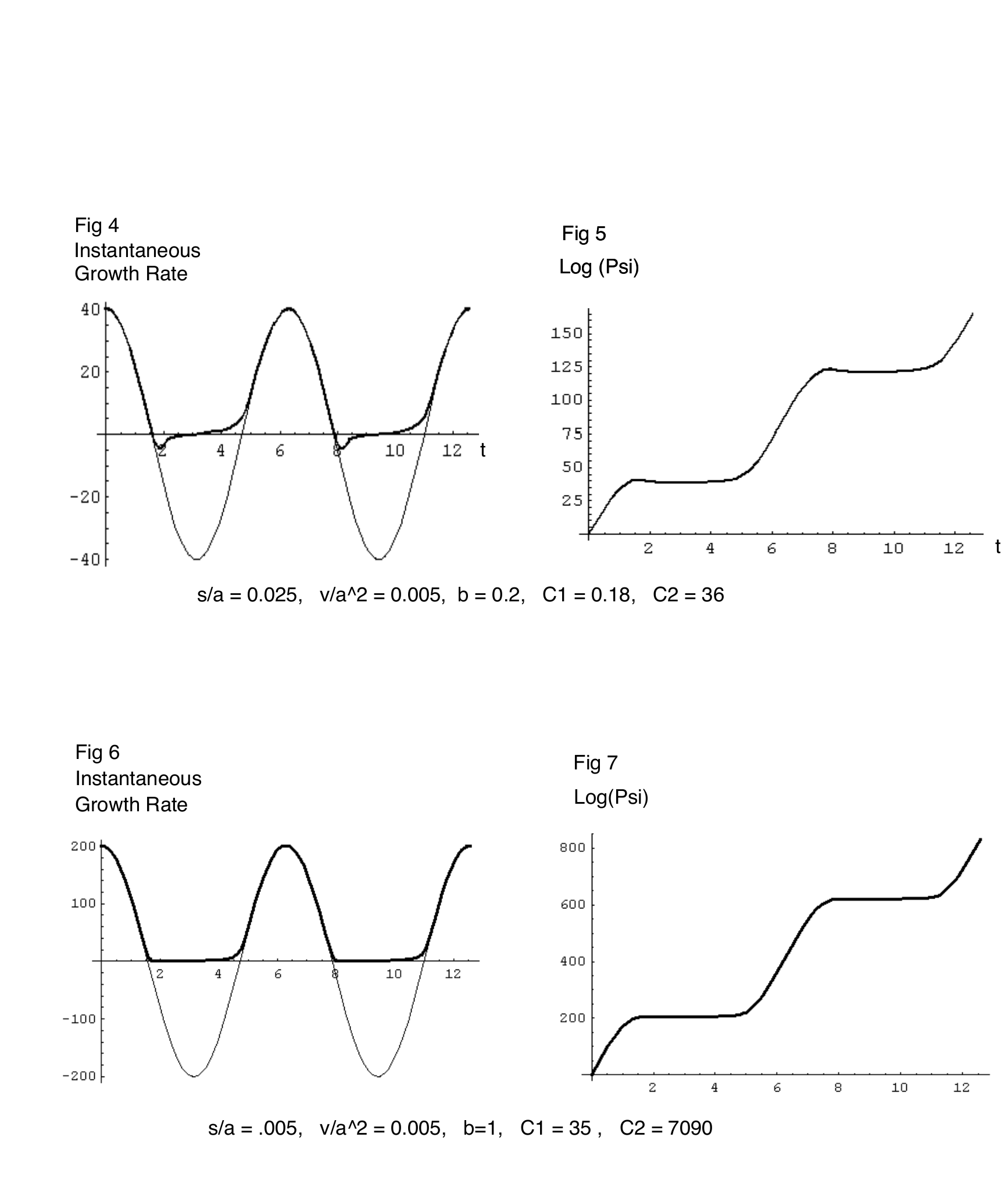}
\end{figure}

\begin{figure}
\centering
\includegraphics[width=\textwidth]{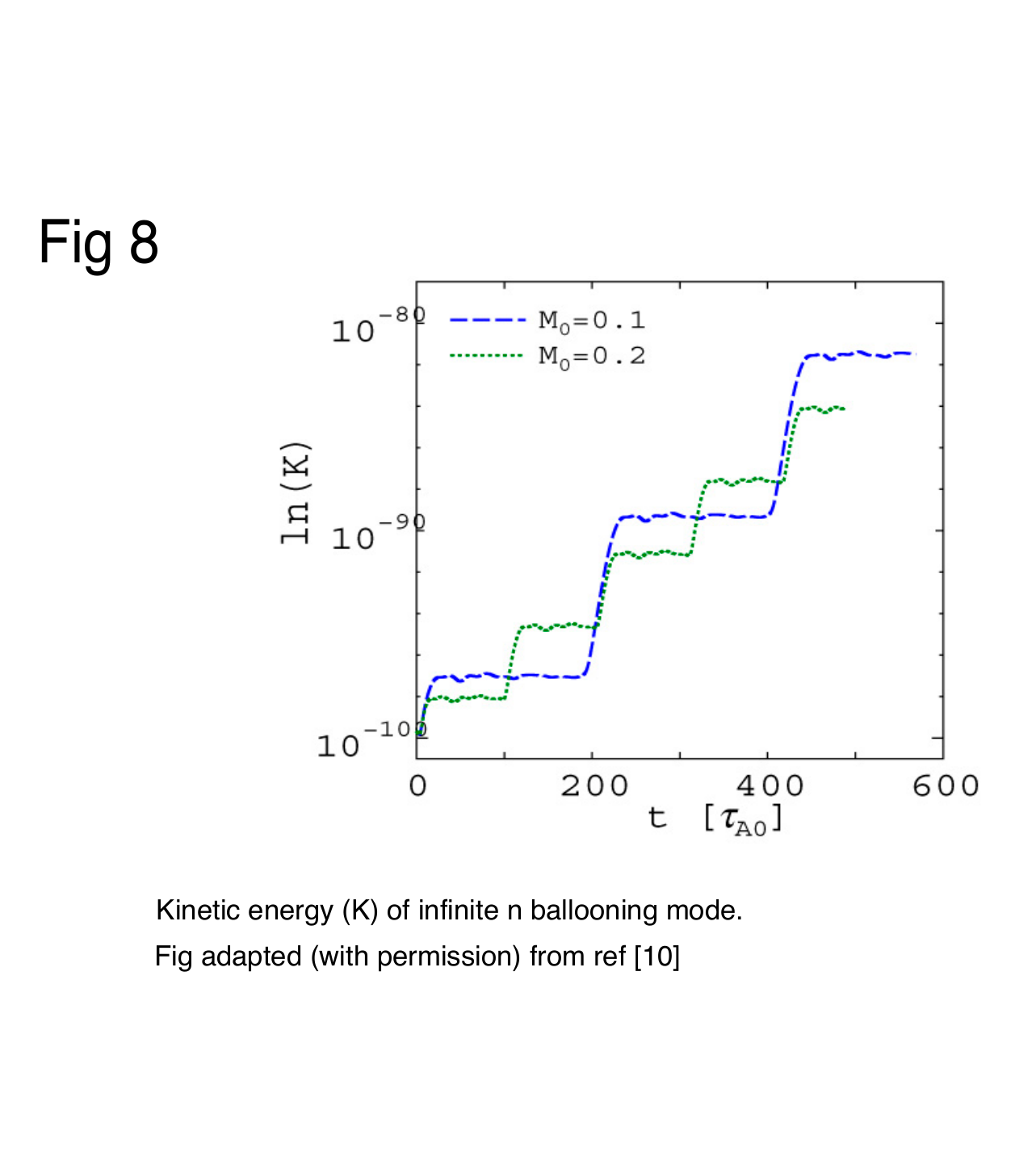}
\end{figure}

\end{document}